\documentclass[useAMS,usegraphicx,usenatbib]{mn2e}
\usepackage{multirow}
\usepackage{rotating}
\usepackage{lscape}
\usepackage{longtable}
\usepackage{ctable}
\usepackage{txfonts}
\usepackage{color}
\title[On the Eccentricity Distribution of Short-Period Single-Planet
Systems]{On the Eccentricity Distribution of Short-Period
Single-Planet Systems}
\author[Ji Wang, Eric B. Ford]{Ji Wang$^{1}$ and Eric B. Ford$^{1}$\thanks{E-mail:
jwang@astro.ufl.edu (JW); eford@astro.ufl.edu
(EBF)}\footnotemark[1]\\
$^{1}$Department of Astronomy, University of Florida, 211 Bryant Space Science Center, Gainesville, FL, USA, 32611}
\begin{document}


\pagerange{\pageref{firstpage}--\pageref{lastpage}} \pubyear{2002}

\maketitle

\label{firstpage}

\begin{abstract}
We apply standard Markov chain Monte Carlo (MCMC) analysis techniques for 50 short-period, single-planet
systems discovered with radial velocity technique. We develop a new
method for accessing the significance of a non-zero orbital eccentricity, namely $\Gamma$ analysis,
which combines frequentist bootstrap approach with Bayesian analysis
of each simulated data set. We find the eccentricity estimations
from $\Gamma$ analysis are generally consistent with results from
both standard MCMC analysis and previous references. The $\Gamma$ method is particular useful for assessing the significance of small eccentricities. Our results
suggest that the current sample size is insufficient to draw robust
conclusions about the roles of tidal interaction and perturbations
in shaping the eccentricity distribution of short-period
single-planet systems. We use a Bayesian population analysis to show
that a mixture of analytical distributions is a good approximation
of the underlying eccentricity distribution. For short-period
planets, we find the most probable values of parameters in the
analytical functions given the observed eccentricities. These
analytical functions can be used in theoretical investigations or as
priors for the eccentricity distribution when analyzing short-period
planets. As the measurement precision improves and sample size
increases, the method can be applied to more complex
parametrizations for the underlying distribution of eccentricity for
extrasolar planetary systems.
\end{abstract}

\begin{keywords}
methods: statistical-planetary systems-techniques: radial velocity.
\end{keywords}

\section{Introduction}

The discovery of exoplanets has significantly advanced our
understanding of formation and evolution of planetary
system~\citep{Wolszczan1994,Mayor1995,Marcy1996}. As of February 2011,
over 500 exoplanets have been discovered including 410 systems
detected by radial velocity (RV)
technique\footnotemark\footnotetext{http://exoplanet.eu/;
http://exoplanets.org/}. The eccentricity distribution of exoplanets
is very different from that of solar system. For sufficiently short-period
planets, it is expected that tidal circularization would lead to
nearly circular orbits. Yet, several short-period planets appear to have
eccentric orbits. Several mechanisms (e.g. planet scattering, Kozai
effect) have been proposed to explain the observed eccentricity
distribution~\citep{Takeda2005,Zhou2007,Ford2008,Juric2008}. This
paper aims to improve our understanding of the true eccentricity
distribution and its implications for orbital evolution.

The Bayesian approach offers a
rigorous basis for determining the posterior eccentricity
distribution for individual system. The Bayesian method is particularly advantageous relative
to traditional bootstrap method when the orbital eccentricity is
poorly constrained by RV data~\citep{Ford2006}. ~\citet{Ford2006}
discussed eccentricity estimation using Markov Chain Monte Carlo
(MCMC) simulation in the framework of Bayesian inference theory and
found a parameter set that accelerates convergence of MCMC for low
eccentricity orbit.  For a population of
planets on nearly circular orbits, eccentricity
estimates for planets on circular orbit are biased resulting in
overestimation of orbital eccentricities~\citep{Zakamska2010}.  
Further complicating matters, the population of known exoplanets is not homogeneous, and the
observed eccentricity distribution is affected by the discovery
method, selection effects and data analysis technique. 

In this paper, we construct a catalog of short-period single-planet
systems using homogeneous RV data reduction process, i.e., standard
MCMC analysis (\S 2.1). In \S 2.2, we describe a new method of
estimating Keplerian orbital parameters, namely $\Gamma$ analysis.
We present the results for
standard MCMC analysis in \S 3. We compare the results of both methods with each other and the
results from previous references. We interpret the results to
investigate how tidal effect and perturbation affect the orbital
eccentricity distribution of short-period planets (\S 4). In \S 5,
we provide an analytical function for the underlying eccentricity
distribution that is able to reproduce the observed the eccentricity
distribution. We discuss the results in \S 6 and summarize our
conclusions in \S 7.

\section{Method}

We select all the systems with: 1) a single known planet discovered
with the radial velocity technique as of April 2010; 2) an orbital
period of less than 50 days; and 3) a publicly available radial
velocity data set. We exclude planets discovered by the transit
technique in order to avoid complications due to selection
effects~\citep{Gaudi2005}. We perform an orbital analysis on each
system in our sample using: 1) a standard MCMC analysis (\S 2.1) and
2) a new method, $\Gamma$ analysis (described in \S 2.2). We focus
on the eccentricity estimation for each planet since the
eccentricity is an important indication of orbital evolution and
tidal interaction.

\subsection{Bayesian Orbital Analysis of Individual Planet}
We performed a Bayesian analysis of the published radial velocity
observation using a model consisting of one low-mass companion
following a Keplerian orbit. If a long-term RV trend is
included in the original paper reporting the RV data or if a linear
trend of more than 1 $\rm{m}\cdot\rm{s}^{-1}\cdot\rm{yr}^{-1}$ is apparent, then we add to the model a constant
long-term acceleration due to distant planetary or stellar
companion.

We calculate a posterior sample using the Markov Chain Monte Carlo
(MCMC) technique as described in ~\citet{Ford2006}. Each state in
the Markov chain is described by the parameter set
$\vec{\theta}=\{P,K,e,\omega,M_0,C_i,D,\sigma_j\}$, where $P$ is
orbital period, $K$ is the velocity semi-amplitude, $e$ is the
orbital eccentricity, $\omega$ is the argument of periastron, $M_0$
is the mean anomaly at chosen epoch $\tau$, $C_i$ is constant
velocity offset (subscript $i$ indicates constant for different
observatory), $D$ is the slope of a long-term linear velocity trend,
and $\sigma_j$ is the ``jitter'' parameter. The jitter parameter
describes any additional noise including both astrophysical noises,
e.g., stellar oscillation, stellar spots ~\citep{Wright2005} and any
instrument noise not accounted for in the reported measurement
uncertainties. The RV perturbation of a host star at time $t_k$ due to a planet on
Keplerian orbit and possible perturbation is given by

\begin{equation}
v_{k,\vec{\theta}}=K\cdot[\cos(\omega+T)+e\cdot
\cos(\omega)]+D\cdot(t_k-\tau),
\end{equation}
where $T$ is the true anomaly which is related to eccentric anomaly
$E$ via the relation,

\begin{equation}
\tan\left(\frac{T}{2}\right)=\sqrt{\frac{1+e}{1-e}}\tan\left(\frac{E}{2}\right).
\end{equation}
The eccentric anomaly is related to the mean anomaly $M$ via Kepler's
equation,

\begin{equation}
E(t)-e\cdot\sin[E(t)]=M(t)-M_0=\frac{2\pi}{P}(t-\tau).
\end{equation}

We choose priors of each parameter as described in
~\citet{Ford2007}. The prior is uniform in logarithm of orbital
period. For $K$ and $\sigma_j$ we use a modified Jefferys prior in
the form of $p(x)\propto(x+x_o)^{-1}$~\citep{Gregory2005} with
$K_{\mathrm min}=\sigma_{j,\mathrm{min}}=0.1$ $\rm{m}\cdot\rm{s}^{-1}$. Priors are uniform
for: $e$ ($0\le e \le 1$), $\omega$ and $M$ ($0\le \omega,M <2\pi$),
$C_i$ and $D$.  We verified that the parameters in $\vec{\theta}$
did not approach the limiting values. We assume each observation
results in a measurement drawn from normal distribution centered at
the true velocity, resulting in a likelihood (i.e., conditional
probability of making the specified measurements given a particular
set of model parameters) of

\begin{equation}
p(\vec{v}|\vec{\theta},M) \propto \prod_k
\frac{\exp[-(v_{k,\vec{\theta}}-v_k)^2/2\sigma^2_k]}{\sigma_k},
\end{equation}
where $v_k$ is radial velocity at time $t_k$, and $v_{k,\theta}$ is
the model velocity at time $t_k$ given the model parameters
$\vec{\theta}$. Noise $\sigma_k$ consists of two parts. One component is from
the observation uncertainty $\sigma_{k,obs}$ reported in the radial
velocity data, and the other is the jitter, $\sigma_j$, which accounts for any unforseen
additional noise including instrument instability and stellar
jitter. The two parts are added in quadrature in order to generate
$\sigma_k$. We calculate the Gelman-Rubin statistic, $\hat{R}$, to
test for nonconvergence of Markov chains.

\begin{figure}
\includegraphics[width=8cm]{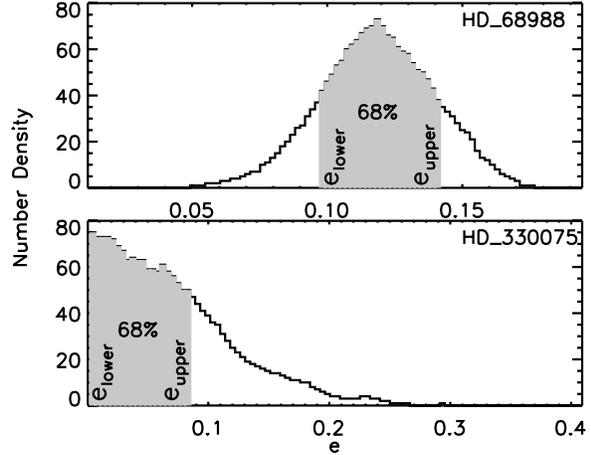}
\caption{Examples of how credible intervals of standard MCMC
analysis are calculated using posterior distribution of $e$. Grey
region contains 68\% of total number of posterior samples of $e$.
\label{fig:MCMC}}
\end{figure}

We perform a MCMC analysis for RV data set of each
system in the sample and obtain posterior samples of $h$ and $k$,
where $h=e\cos{\omega}$ and $k=e\sin{\omega}$. This parameterization
has been shown to be more effective in description of the
eccentricity distribution for low eccentric orbits~\citep{Ford2006}. We take steps in $h$ and $k$ and adjust the acceptance rate according
to the Jacobian of the coordiante transformation, so as to maintain a
prior that is uniform in $e$ and $\omega$.
Mean values, $\bar h$ and $\bar k$, from posterior samples of $h$
and $k$ are adopted to calculate $e_{MCMC}$ using the equation
$e_{MCMC}=\sqrt{\bar h^2+\bar k^2}$. The posterior distribution of
$e$ is not always Gaussian distribution especially near $e\sim$0.
Therefore, it is not appropriate to calculate the uncertainty of $e$
using the equation of error propagation in which gaussian noise is
assumed. We use posterior distribution of $e$ to infer the credible
interval of $e$. The boundaries of the region where 68\% posterior
samples populate are adopted as $e_{lower}$ and
$e_{upper}$ (Appendix \ref{app:comp}). Fig. \ref{fig:MCMC}
illustrates two examples of how the credible intervals are inferred
for HD 68988 (eccentric orbit) and HD 330075 (circular orbit).

\subsection{$\Gamma$ Analysis of Individual Systems}

A fully Bayesian analysis of the population of exoplanet
eccentricities would be computationally prohibitive due to the large number of dimensions. Therefore, we
develop a hybrid Bayesian-frequentist method to evaluate the significance of a non-zero eccentricity
measurement. We combine a bootstrap style approach of generating and analyzing synthetic data sets with MCMC
analysis of each synthetic data set to obtain a frequentist
confidence level for each eccentricity that accounts for biases. First, we perform the
standard MCMC analysis described in \S 2.1 on the real RV data set and adopt the mean value
of each orbital parameter in $\vec{\theta}$ except $e$. We generate a series of
simulated radial velocity data sets at different values of $e$. The
adopted $K$ is scaled accordingly to $K\propto(1-e^2)^{-0.5}$. The
simulated radial velocity data has the same number of observations,
and each simulated observation takes place at exactly the same time
and the same mean anomaly as the real observation. Gaussian noise
with standard deviation of $\sigma_k$ (\S2.1) are added to simulated
radial velocity data sets at different eccentricities. Each
simulated RV data set has the same reported RV
measurement uncertainties as the real RV observations. Standard MCMC analysis is then performed on each of the simulated RV data sets.

\begin{figure}
\includegraphics[width=8cm]{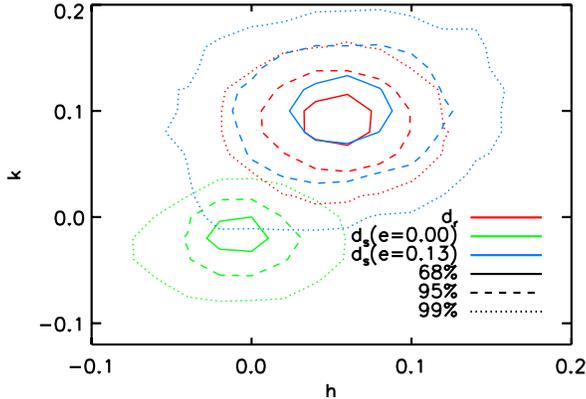}
\caption{Contours of posterior distribution in $h$ and $k$ space for
HD 68988 (Solid-68\% of sample points included; dashed-95\% of
sample points included; dotted-99\% of sample points included). Red
contours are posterior distribution for real observation, green
contours are for simulated RV data set with $e$=0.00, and blue
contours are for simulated RV data set with $e$=0.13. Eccentricity
of HD 68988 is 0.1250$\pm$0.0087 according to ~\citet{Butler2006}.
\label{fig:Gamma_illus}}
\end{figure}

For both real and simulated data sets, we construct a
two-dimensional histogram using the posterior samples in $(h,k)$
space to approximate a two-dimensional posterior distribution for
$h$ and $k$, $d_r(h_i,k_j)$ and $d_s(h_i,k_j)$, where $i$ and $j$
denote bin indices, and the subscripts $r$ and $s$ denote the real
and simulated data set. We compare the distribution for each
simulated data set $d_s(h_i,k_j)$ to the distribution for real
radial velocity data set $d_r(h_i,k_j)$. To quantify the similarity
between $d_s(h_i,k_j)$ and $d_r(h_i,k_j)$, we calculate the
statistic defined as
$\Gamma=[{\displaystyle\sum_{i=1}^{N}\displaystyle\sum_{j=1}^{N}
(d_s(h_i,k_j)-\mu_s)\cdot(d_r(h_i,k_j)-\mu_r)}]/[{\sigma_s\sigma_r(N^2-1)}]$,
where $N$ is the number of bins in $h$ or $k$ dimension, $\mu$ and
$\sigma$ represent mean and standard deviation. In other
words, the $\Gamma$ statistic is obtained by cross-correlating two
posterior distributions in $h$ and $k$ space. Fig
\ref{fig:Gamma_illus} illustrates the process by which we obtain
$\Gamma$ for the case of HD 68988. If $d_s(h_i,k_j)$ matches
$d_r(h_i,k_j)$, we expect to obtain a $\Gamma$ value that approaches
unity (blue and red contours).  If the samples differ significantly then $\Gamma$ decreases towards zero (red and green contours).
For each eccentricity, we simulated 21 radial velocity data sets and
compare $d_s(h_i,k_j)$ with $d_r(h_i,k_j)$ to obtain 21 $\Gamma$
statistics between simulated and real RV data. We
choose the median value $\bar\Gamma$ as an indicator of overall
similarity at given eccentricity.

\begin{figure}
\includegraphics[width=8cm]{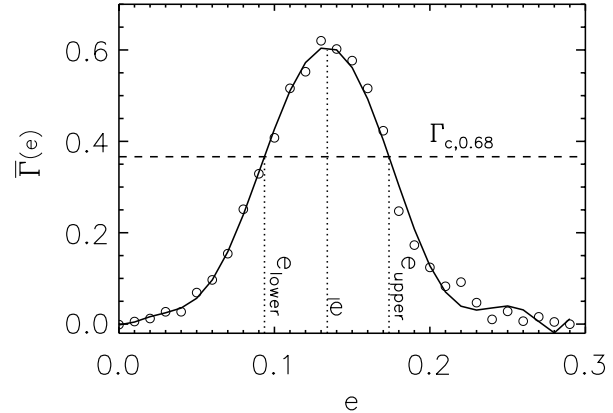}
\caption{$\bar\Gamma$ as a function of eccentricity $e$  for HD
68988. Open circles are results from simulations, solid line is the
result of polynomial fitting. The long dashed line is the critical
threshold, $\Gamma_{c,0.68}$ at the 68\% confidence level.
\label{fig:hd68988}}
\end{figure}

Based on above analysis of simulated radial velocity data with
different input eccentricities $e$, we obtain a relationship between
$\bar\Gamma$ and $e$, i.e. $\bar\Gamma(e)$. We use a high-order
polynomial to interpolate for $\bar\Gamma(e)$. We define $\bar e$ to
be the eccentricity at which $\bar\Gamma(e)$ reaches maximum and we
interpret $\bar e$ as an estimator of eccentricity. For HD 68988
(Fig. \ref{fig:hd68988}), input eccentricity ranges from 0.00 to
0.29 with step size of 0.01. $\bar\Gamma(e)$ reaches maximum at
$e=0.134$. We estimate statistical confidence level of $\bar e$
using every pair of posterior eccentricity samples calculated from
the data sets that are generated assuming the same eccentricity.
Consider the example of HD 68988 again: 21 sets of
posterior distribution in $h$ and $k$ space, $d_s(h_i,k_j)$, are
obtained. Comparison between each pair gives a $\Gamma$ statistic
between simulated RV data. $\Sigma_{i=1}^{20}=210$ $\Gamma$ statistics in
total for simulated RV data sets are calculated at eccentricity of 0.13 and 68.1\% of pairs
(143 out of 210) have a $\Gamma$ statistic greater than 0.3714. We
define this value, $\Gamma_{c,0.68}$, as the critical $\Gamma$ value
for HD 68988 at 68\% confidence level for $e=0.13$. Therefore, if
$\Gamma$ statistic obtained in comparison between $d_s(h_i,k_j)$ and
$d_r(h_i,k_j)$ is less than $\Gamma_{c,0.68}$, we argue that the
eccentricity inferred from simulated RV data set is not consistent
with the observed eccentricity of the system at 68\% confidence
level. In the case where $\bar{e}$ is located between grids of
simulated e values, we calculate $\Gamma_{c,0.68}$ at $\bar e$ using
interpolation of $\Gamma_{c,0.68}$ at nearby e values. We use a
high-order polynomial to approximate the discrete data
$\bar\Gamma(e)$. The polynomial is later used to infer $\bar{e}$,
lower and upper limit of eccentricity. For HD 68988,
$\Gamma_{c,0.68}$ is 0.3663 at $\bar{e}=0.134$ after interpolation.
Using the relationship between $\bar\Gamma$ and $e$ (Fig.
\ref{fig:hd68988}), we look for the $e$ values corresponding to
$\Gamma_{c,0.68}$ as estimators of the lower and upper limit for
eccentricity of the planet system at a 68\% confidence level. In HD
68988, we obtained $e=0.134^{+0.040}_{-0.040}$ using $\Gamma$ analysis. In
comparison, we have obtained $e=0.119^{+0.025}_{-0.022}$ using a standard MCMC
analysis and ~\citet{Butler2006} reported $e=0.125\pm0.009$.

\section{Results for Individual Planets}
In our sample of 50 short-period single-planet systems, we
successfully analyzed 42 systems using $\Gamma$ analysis, and 46
systems using standard MCMC analysis (Appendix \ref{app:comp}). All
the error are based on a 68\% confidence level ($\Gamma$ analysis)
or a 68\% credible interval (MCMC). The unsuccessful cases in
standard MCMC analysis are HD 189733, HD 219828, HD 102195 and GJ
176. In HD 189733, most of the RV data points were taken during
observation of Rossiter-McLaughlin effect, which is not modeled
here. MCMC analysis fails for HD 219828 and GJ 176 because of low
signal to noise ratio (K=7 $\rm{m}\cdot\rm{s}^{-1}$ and $n_{obs}$=20 for HD 219828 and
K=4.1 $\rm{m}\cdot\rm{s}^{-1}$ and $n_{obs}$=57 for GJ 176). RV data points of HD 102195
were taken at 3 observatories and MCMC analysis is complicated using
different observatory offsets. In addition to the above, $\Gamma$
analysis was unsuccessful for HD 162020, GJ 86, HD 17156 and HD
6434. Since $\Gamma$ analysis involves generating simulated RV data,
the limited number of observations and partial phase coverage for
these systems can cause poor convergence for some simulated data
sets. These limitations become more severe for systems with high
eccentricity (e.g. HD 17156, $e=0.684$) since phase coverage is more
important for eccentric orbits.

\begin{figure}
\begin{center}
\includegraphics[width=8cm]{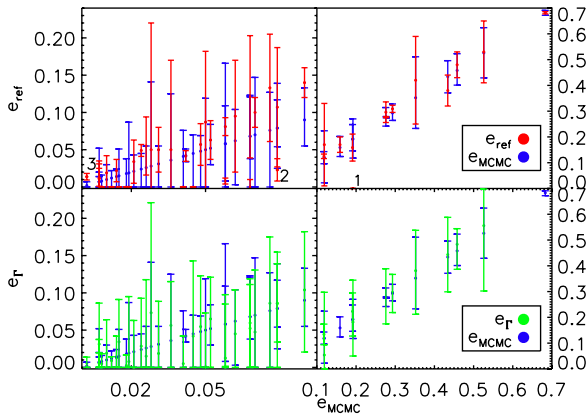}
\caption{Top: comparison between standard MCMC analysis (blue) and
previous references (red); Bottom: comparison between standard MCMC
analysis (blue) and $\Gamma$ analysis (green). The systems where
disagreements take place are marked with a number: 1, HD 149026; 2,
$\tau$ Boo; 3, HD 195019. \label{fig:comp}}
\end{center}
\end{figure}

\subsection{Comparison: Standard MCMC and References}
\label{secCompMcmcRef}
In Fig. \ref{fig:comp} (top), we compare results from two sources,
standard MCMC analysis and previous references. In most cases the two
methods provide similar results.  We found there are 3
systems for which the eccentricity estimates are not consistent,
i.e., the published eccentricity error bar does not overlap the 68\% credible interval from our MCMC analysis.
They are HD 149026, $\tau$ Boo, HD
195019. For HD 149026, ~\citet{Sato2005} set the eccentricity to be
zero when fitting the orbit. In contrast, we treat eccentricity as a
variable and the standard MCMC method found that the 68\% credible
interval for the systems mentioned above does not include zero. In
addition, HD 149026 b is a known transiting planet and transit
photometry provides additional constraints on eccentricity which
we have not included~\citep{Charbonneau2003}. ~\citet{Knutson2009}
measured $\Delta t_\Pi=20.9^{+7.2}_{-6.2}$ minute(2.9$\sigma$) for
HD 149026 which is inconsistent with zero eccentricity, because
$e\ge e\cos\omega\simeq\frac{\pi}{2P}\Delta t_{\Pi}$, where $P$ is
period and $\Delta t_\Pi$ is the deviation of secondary eclipse from
midpoint of primary transits. Standard MCMC results for other two
systems ($\tau$ Boo and HD 195019) are not consistent with those from previous references even
though eccentricity was treated as variable in previous references.  
\citet{Butler2006} report $e=0.023\pm0.015$ for $\tau$ Boo and $e=0.014\pm0.004$ for HD 195019. On
the contrary, standard MCMC analysis gives
$e$=0.0787$^{+0.0382}_{-0.0246}$ for $\tau$ Boo and
$e$=0.0017$^{+0.0049}_{-0.0017}$ for HD 195019 (See Appendix
\ref{app:comp}).  

\subsection{Comparison: Standard MCMC and $\Gamma$ Analysis}

Fig. \ref{fig:comp} (bottom) compares the results from standard MCMC
analysis and $\Gamma$ analysis. We find that the 68\% credible/confidence intervals for the two methods overlap
in the cases where there are discrepancies between standard MCMC analysis and previous references (see \S~\ref{secCompMcmcRef}). The confidence interval from the $\Gamma$
analysis is generally larger than the credible interval from a standard MCMC analysis.
The larger uncertainty from $\Gamma$ analysis is likely due to the analysis accounting for the uncertainty in each velocity observation twice, first when
generating synthetic data sets and a second time when analyzing the simulated data.  Thus, the $\Gamma$ analysis results in slightly larger uncertainty in eccentricity estimation. 

In order to understand the behavior of standard MCMC
analysis and $\Gamma$ analysis for planets on nearly circular orbits, we conduct an additional experiment generating many synthetic data sets where each system has a single planet on a circular orbit.  We assume that they are observed at the same times and with the same RV measurement precisions as actual RV data sets.  In order to understand the bias of each method for nearly circular systems, we compare the output eccentricities and their uncertainties. Using standard MCMC analysis, we find that $76.4\pm{2.9}\%$ of the simulated data sets are consistent with zero using a 68\% credible interval, and $23.6\pm{1.6}\%$ of the simulated data sets have 68\% credible intervals that do not include zero.  In contrast, for $16.8\pm{1.4}\%$ of the simulated data sets, the $\Gamma$ analysis does not result in a 68\% confidence interval that includes zero.  In both cases, more than 68\% of data sets are consistent with a circular orbit at a 68\% level using either method.  Using the $\Gamma$ analysis, $6.8\pm{3.0}\%$ more simulated data sets are consistent with a circular orbit than based on the standard MCMC analysis.  This confirms our intuition that the $\Gamma$ analysis is a less biased method for analyzing systems at very small eccentricity.  Thus, the $\Gamma$ analysis may be a useful tool in assessing the significance of a measurement of a small non-zero eccentricity.  In particular, we find 5 cases (11.4\%) in which $e_{lower}$=0 for $\Gamma$ analysis even though $e_{lower}$ for MCMC is greater than zero (e.g., HD 46375, HD 76700, HD 7924, HD 168746, HD 102117). 

A similar experiment is conducted except that an eccentricity of 0.2 is assigned to each system instead of zero eccentricity.  Again, we assess the accuracy of the two methods by comparing the input and output eccentricities.  When using the MCMC method, we find that the 68\% credible interval for the eccentricity does not include the input eccentricity for $26.0\pm2.1\%$ of simulated data sets.  When using the $\Gamma$ method, we find that the 68\% confidence interval for the eccentricity does not include the input eccentricity for $18.9\pm1.8\%$ of simulated data sets.  Again, there is a larger fraction of results from the MCMC method that are not consistent with the input at a sizable eccentricity, indicating $\Gamma$ analysis is less likely to reject the correct eccentricity.  We also investigate the bias of the two methods at a significant eccentricity (i.e. $e$=0.2). In the cases where the output eccentricities are consistent with the input, we find that $47.9\pm{3.3}\%$ of the output eccentricities are below 0.2 while $52.1\pm{3.5}\%$ of outputs are over 0.2 for MCMC method indicating the MCMC method is not biased at a sizable eccentricity, which agrees with the finding from~\citet{Zakamska2010}. In comparison, $\Gamma$ analysis is also an unbiased analysis with $49.9\pm{3.3}\%$ below input and $50.1\pm{3.3}\%$ exceeding input. Therefore, we find no evidence for significant bias of either method for data sets with a significant eccentricity.

\subsection{Discussion of $\Gamma$ Analysis}

The different methods for analyzing Doppler observations are complimentary.
Bayesian methods and MCMC in particular are routinely used to sample
from the posterior distribution for the Keplerian orbital parameters
for a given system.  However, the analysis of an exoplanet population
is more complicated than simply performing a Bayesian analysis of each
system.  To illustrate this point, consider a population of planets
that all have exactly circular orbits.  Due to measurement
uncertainties and finite sampling, the ``best-fit'' eccentricity for
each system will be non-zero.  Similarly, since eccentricity is a
positive-definite quantity, the analysis of each system will result in
a posterior distribution that has significant support for $e>0$.  This
property remains even if one combines many point estimates (e.g.,
``best-fit''), frequentist confidence intervals or Bayesian posterior
distributions.  While the posterior distribution for the orbital
parameters represents the best possible analysis of an individual
system, the inevitable bias for nearly circular orbits is a potential
concern for population analyses.  Therefore, it is important to apply
different methods for population analyses (e.g., ~\citet{Hogg2010, Zakamska2010}).

Since we intend to investigate the potential role of tidal effects on
the eccentricity distribution of short-period planets, we developed a
hybrid technique to assess the sensitivity of our results to bias in
the posterior distribution for planets with nearly circular orbits.
This hybrid technique ($\Gamma$ analysis) involves performing Bayesian
analyses of each individual planetary system along with several
simulated data sets, each of varying eccentricity.  The MCMC analysis
of each data set allows us to account for the varying precision of
eccentricity measurements depending on the velocity amplitude,
measurement precision, number of observations and phase coverage.  We
assess the extent of the eccentricity biases by performing the same
analysis on simulated data sets with known eccentricity.  We compare
the posterior densities for the actual data set to the posterior
density for each of the simulated data sets to determine which input
model parameters are consistent with the observations.  We can
construct frequentist confidence intervals based on Monte Carlo
simulations (i.e., comparing the posterior distributions for the
synthetic data sets to each other).  

The basic approach of the $\Gamma$ analysis is similar to likelihood-free methods more commonly used in approximate Bayesian computation.  In this case, we do have a likelihood which allows us to sample from the posterior probability distribution using standard MCMC.  We compare the posterior densities calculated for several simulated data sets to the posterior density of the actual observations, so as to assess the accuracy and bias of the standard MCMC analysis.


\begin{figure}
\begin{center}
\includegraphics[width=8cm]{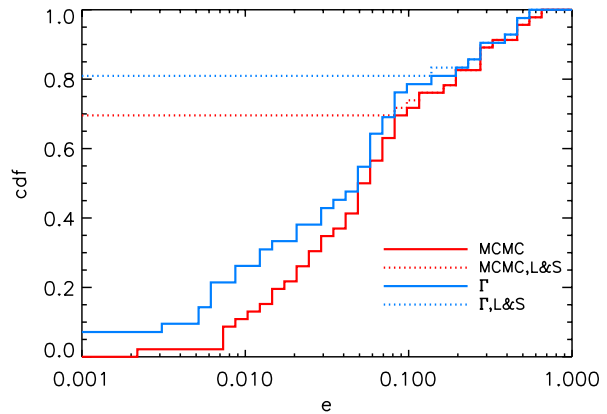}
\caption{Cumulative distributions functions (CDFs) of eccentricities
from different methods. The solid red line is for $e_{MCMC}$, adopted from
MCMC method (Appendix \ref{app:comp}).  The dotted red line is similar to MCMC,
but any eccentricity with $e_{lower}$ of 0 for a 95\% credible interval is
assigned to 0.  The blue lines are for $\Gamma$ analysis, where the
solid line is for $e_\Gamma$ from the $\Gamma$ analysis (Appendix \ref{app:comp})
and the dotted line is similar, but any eccentricity with $e_{lower}$ of 0
for a 95\% confidence interval is assigned to 0.
 \label{fig:cdf_ls}}
\end{center}
\end{figure}

The problem of biased eccentricity estimators for nearly circular orbits is familiar from previous studies of binary stars.  In particular, \citet{Lucy1971} investigated the possibility of mistakenly assigning an eccentric orbit to a circular spectroscopic binary due to inevitable measurement errors.  As many spectroscopic binaries may have been affected by tidal circularization, they suggested assigning a circular orbit to any system for which the eccentricity credible interval contained 0.  When studying a population of systems for which circular orbits are common, this approach significantly reduces the chance of erroneously concluding the system has a non-zero eccentricity.  One obvious disadvantage of this approach is that it would lead to a negative bias for systems where the eccentricity is of order $\sigma/K$, where $\sigma$ is the typical measurement precision and $K$ is the velocity amplitude.  For binary stars, $\sigma/K$ may be small enough that this is not a significant concern.  For exoplanets, where $\sigma/K$ may be as small as $\sim~2-3$, such a procedure would result in a significant negative bias for many systems.  The $\Gamma$ analysis offers an alternative approach, which may be particularly useful when  analyzing the eccentricity distribution of a population of planetary systems.  

For the sake of comparison, we consider a modernized version of the ~\citet{Lucy1971} approach which is based on the posterior distribution from a standard MCMC analysis or the confidence interval from our $\Gamma$ analysis.  We construct a histogram or cumulative distribution of the eccentricities for a population of systems, using a single summary statistic for each system:  the posterior mean for the standard MCMC analysis or the eccentricity that maximizes the $\Gamma$ statistic.  Following ~\citet{Lucy1971}, we adopt an eccentricity of zero for any system for which the 95\% significance level ($\Gamma$ method) or the 95\% credible interval (MCMC) includes $e=0$. The cumulative distribution functions of the eccentricities using different methods are plotted in Fig. \ref{fig:cdf_ls}. Based on the generalized ~\citet{Lucy1971} approach, $\sim$81\% (70\%) of the short-period planet systems in our sample are consistent with circular orbits using the $\Gamma$ analysis (standard MCMC analysis).  Clearly, the ~\citet{Lucy1971} approach results in a large fraction being assigned a circular orbit, largely due to the choice of a 95\% threshold.  The fraction assigned a circular orbit is sensitive to the size of the credible interval used when deciding whether to set each eccentricity to zero.  There is no strong justification for the choice of the 95\% threshold (as opposed to 68\% or 99.9\% threshold) and tuning the threshold to agree with other methods negates the primary advantage of the ~\citet{Lucy1971} method, that it requires no additional computations.  Therefore, we do not recommend using the ~\citet{Lucy1971} approach to learn about the eccentricity distribution for a population when $\sigma/K$ is not large.

\section{Tidal Interaction Between Star and Planet}

Several factors affect the eccentricity distribution of short-period
planets including tidal interaction between host star and planet and
possible perturbation of an undetected companion. We will discuss
how these factors affect the eccentricity distribution and whether
the effect is observable.

\begin{figure}
\begin{center}
\includegraphics[width=8cm]{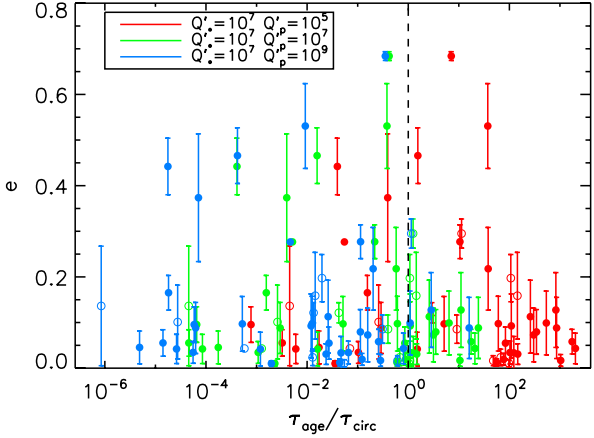}
\caption{Distribution of short-period single-planet systems in
($e$,$\tau_{age}/\tau_{circ}$) space. Filled circles are systems
showing no linear RV trend and open circles are systems showing
long-term linear RV trends. Different colors indicate different combinations of $Q^\prime_\ast$ and $Q^\prime_p$ \label{fig:eCircAge}}
\end{center}
\end{figure}

In order to understand the influence of tidal interaction on
eccentricity distribution,  we first divide our sample into two
subsets, one subset contains systems with a long-term RV trend while
the other subset contains systems that do not show a long-term RV
trend (Fig. \ref{fig:eCircAge}). The systems that are noted with a
long-term velocity trend include 51 PEG, BD -10 3166, GJ 436, GJ 86,
HD 107148, HD 118203, HD 149143, HD 68988, HD 7924, HD 99492 and
$\tau$ Boo. We further divide the no-trend subset into two groups,
one group is distinguished by $\tau_{age}/\tau_{circ}\ge1$, and the
other group is distinguished by $\tau_{age}/\tau_{circ}<1$, where
$\tau_{circ}$ is tidal circularization time scale and $\tau_{age}$
is the age of the host star. We investigate whether there is a
significant difference in the eccentricity distribution between
these two groups as expected if tidal interaction is an important
factor in shaping eccentricity distribution.
Following~\citet{Matsumura2008}, we estimate $\tau_{circ}$ using:

\begin{equation}
\tau_{circ}=\frac{2}{81}\frac{Q^\prime_p}{n}\frac{M_p}{M_\ast}\bigg(\frac{a}{R_p}\bigg)^5
\bigg[\frac{Q^\prime_p}{Q^\prime_\ast}\bigg(\frac{M_p}{M_\ast}\bigg)^2\bigg(\frac{R_\ast}{R_p}\bigg)^5F_\ast+F_p\bigg]^{-1},
\end{equation}
where the subscripts $p$ and $\ast$ denote planet and star, $M$ is
mass, $R$ is radius, $a$ is semi-major axis, $Q^\prime$ is modified
tidal quality factor and $n=[G(M_\ast+M_p)/a^3]^{1/2}$ is the mean
motion. ~\citet{Matsumura2008} adopted $10^6$ as a typical value for
$Q^\prime_\ast$ for short period planetary system host stars and
considered $Q^\prime_p$ ranging from $10^5$ to $10^9$. We use
$Q^\prime_\ast=[10^6,10^7]$ and $Q^\prime_p=[10^5,10^7,10^9]$ in our
analysis. The factors $F_\ast$ and $F_p$ are defined in the
following two equations:

\begin{equation}
F_\ast=f_1(e^2)-\frac{11}{18}f_2(e^2)\frac{\Omega_{\ast,rot}}{n},
\end{equation}

\begin{equation}
F_p=f_1(e^2)-\frac{11}{18}f_2(e^2)\frac{\Omega_{p,rot}}{n},
\end{equation}
where $\Omega$ is rotational frequency. For short-period planets one could set
$\Omega_{p,rot}/n=1$ based on the assumption that all the planets in
our sample have been synchronized since
$\tau_{synch}\sim10^{-3}\tau_{circ}$~\citep{Rasio1996}.
In order to check whether our conclusion
is sensitive to the choice of $\Omega_{\ast,rot}/n$, we conduct
calculations with other $\Omega_{\ast,rot}/n$ values in which we
choose stellar rotation period  to be 3, 30, and 70 days for all the
stars. We find that this range for $\Omega_{\ast,rot}/n$ does not
change the conclusions in the paper. Therefore, for future discussion, we adopt $\Omega_{\ast,rot}/n=0.67$, which results in stellar rotation periods consistent with typical values from 3 to 70
days~\citep{Matsumura2008}. The uncertainties in $\Omega_{\ast,rot}/n$ are accounted for
by our subsequent data analysis (Equation. \ref{eq:specialproc}). And $f_1$ and $f_2$ are approximated by the equations:

\begin{equation}
f_1(e^2)=(1+\frac{15}{4}e^2+\frac{15}{8}e^4+\frac{5}{64}e^6)/(1-e^2)^{13/2},
\end{equation}

\begin{equation}
f_2(e^2)=(1+\frac{3}{2}e^2+\frac{1}{8}e^4)/(1-e^2)^{5}.
\end{equation}

Planet radius $R_p$ is estimated based on ~\cite{Fortney2007}. We assume that the planet and host star are formed at the same
epoch. We assume that planet structure
is similar to Jupiter with a core mass fraction of $25
M_\oplus/M_J=7.86\%$. Radii of GJ 436 b and HD 149026 b are adopted
from reference papers because there is a factor of $>$2 difference
between observed values~\citep{Torres2008} and theoretically
calculated values. Stellar radius and age estimations are obtained
from the following sources with descending priority: 1,
~\citet{Takeda2007}; 2, $nsted.ipac.caltech.edu$; 3, $exoplanet.eu$.
The calculated $\tau_{circ}$ values are presented in
Appendix \ref{app:cat} in addition to the results of MCMC analysis
of individual system and other stellar and planetary properties.

\begin{figure}
\begin{center}
\includegraphics[width=8cm]{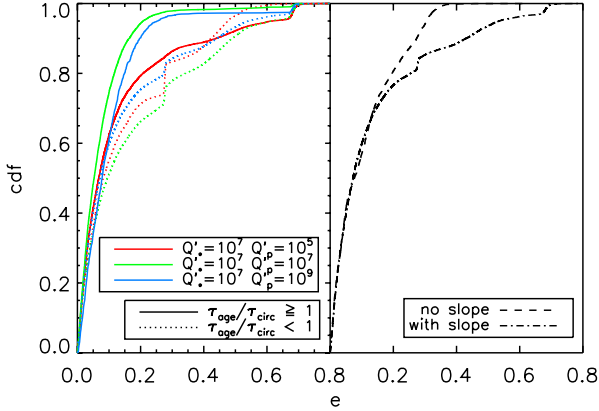}
\caption{The left panel compares the cumulative distribution
function of eccentricity for two groups of planets: 1)
$\tau_{age}/\tau_{circ}\ge1$ (solid line) and 2)
$\tau_{age}/\tau_{circ}<1$ (dotted line). The right panel compares
cumulative distribution functions of eccentricity for two subsets of
planets: planets without a long-term RV slope (dashed line) and
planets with a long-term RV slope (dash-dotted line).
\label{fig:cdfe}}
\end{center}
\end{figure}

We use the eccentricity posterior samples for each system for which
the standard MCMC analysis was successful (i.e., results of \S2.1)
to construct the eccentricity samples of two groups separated by
$\tau_{age}/\tau_{circ}$. We note that there are considerable
uncertainties in the estimation of $\tau_{age}$ and $\tau_{circ}$,
so $\tau_{age}/\tau_{circ}>1$ does not necessarily mean that the
actual system age is larger than the actual circularization time. We
consider the sensitivity of our results to these uncertainties by
adopting a probability function:

\begin{equation}
 \rho(\frac{\tau_{age}}{\tau_{circ}})= \left\{ \begin{array}{ll}
 1-0.5\exp\bigg[-\eta\cdot(\frac{\tau_{age}}{\tau_{circ}}-1)\bigg] &\mbox{if $\frac{\tau_{age}}{\tau_{circ}}\ge1$} \\
0.5\exp\bigg[-\eta\cdot(\frac{\tau_{circ}}{\tau_{age}}-1)\bigg]
&\mbox{if $\frac{\tau_{age}}{\tau_{circ}}<1$}
       \end{array} \right.
\label{eq:specialproc}
\end{equation}
where $\eta$ is a parameter tuning the confidence of $\tau_{age}$
and $\tau_{circ}$ estimation. For example, if
$\tau_{age}/\tau_{circ}=2$ and $\eta=1$, then
$\rho(\tau_{age}/\tau_{circ})=0.816$, meaning there is 81.6\% chance
that the system is from the group of $\tau_{age}/\tau_{circ}\ge1$
because of the uncertainties in $\tau_{age}$ and $\tau_{circ}$
estimation. Therefore, we take 81.6\% of the eccentricity posterior
samples of the system to construct eccentricity sample for the group
of $\tau_{age}/\tau_{circ}\ge1$ and the remaining 18.4\%
eccentricity posterior samples to construct eccentricity sample for
group of $\tau_{age}/\tau_{circ}<1$. The $\eta$ parameter reflects
our confidence in $\tau_{age}$ and $\tau_{circ}$ estimation. If we
are not very confident in the estimation of $\tau_{age}$ and
$\tau_{circ}$, then we set $\eta$ to a small value approaching zero,
so half of the eccentricity posterior samples for each system are
assigned to the group with $\tau_{age}/\tau_{circ}\ge1$ and the the
other half are assigned to the group with
$\tau_{age}/\tau_{circ}<1$. After constructing the eccentricity
sample for the two groups, we use two-sample K-S test to test the
null hypothesis that these two samples from two groups were drawn
from the same parent distribution. The results (Table
\ref{tab:KStest}) show that we are unable to exclude the null
hypnosis at a low $p$ value (statistic of two-sample K-S
test) because of the small effective sample size
($N^\prime$$\sim$8). If $\Delta_{max}=0.2$, where $\Delta_{max}$ is the maximum
difference between cumulative distribution functions of two groups,
we can exclude the null hypnosis at $p=0.05$ only if $N^\prime$ is
more than 44. In comparison, our current sample size is inadequate
to draw a statistically significant conclusion on whether or not the
the groups are from the same parent distribution. However, we do see
a hint of a difference between cumulative distribution functions of
two groups (Fig. \ref{fig:cdfe} left), there are more systems with
low-eccentricity for the group with $\tau_{age}/\tau_{circ}<1$,
which is a consequence of tidal circularization. We also find that
the conclusion is unchanged for a wide range of $\eta$,
$Q^\prime_\ast$, and $Q^\prime_p$ values.

\begin{table}

\caption{Two-sample K-S test result. $\Delta_{max}$ is the maximum
difference between cumulative distribution functions of eccentricity
for two groups separated by $\tau_{age}/\tau_{circ}=1$. $N^\prime$
is the effective sample size, calculated by $(N_1\cdot
N_2)/(N_1+N_2)$, p is the significance level at which two-sample K-S
test rejects the null hypothesis that the two eccentricity samples
are from the same parent distribution; $\eta$ is a parameter tuning
the confidence of $\tau_{age}$ and $\tau_{circ}$ estimation.
\label{tab:KStest}}
\begin{tabular}{|p{1.2cm}|p{1.2cm}|p{0.8cm}|p{0.8cm}|p{0.8cm}|p{0.8cm}|}
\hline \multicolumn{3}{|c|}{$Q^\prime_p$} & $10^5$ & $10^7$ & $10^9$ \\
\hline

\multirow{9}{*}{$Q^\prime_\ast=10^6$} & \multirow{3}{*}{$\eta$=1000} & $\Delta_{max}$ & 0.088 & 0.175 & 0.086 \\
\cline{3-6}
 & & $N^\prime$ & 7.57 & 8.76 & 7.57 \\ \cline{3-6}
 & & p & 1.00 & 0.93 & 1.00 \\ \cline{2-6}
 & \multirow{3}{*}{$\eta$=1} & $\Delta_{max}$ & 0.103 & 0.160 & 0.093 \\ \cline{3-6}
 & & $N^\prime$ & 7.72 & 8.72 & 7.02 \\ \cline{3-6}
 & & p & 1.00 & 0.97 & 1.00 \\ \cline{2-6}
 & \multirow{3}{*}{$\eta$=0.001} & $\Delta_{max}$ & 0.043 & 0.027 & 0.061 \\ \cline{3-6}
 & & $N^\prime$ & 8.59 & 8.50 & 8.08 \\ \cline{3-6}
 & & p & 1.00 & 1.00 & 1.00 \\ \hline


\multirow{9}{*}{$Q^\prime_\ast=10^7$} & \multirow{3}{*}{$\eta$=1000} & $\Delta_{max}$ & 0.088 & 0.286 & 0.190 \\
\cline{3-6}
 & & $N^\prime$ & 7.57 & 8.42 & 2.80 \\ \cline{3-6}
 & & p & 1.00 & 0.43 & 1.00 \\ \cline{2-6}
 & \multirow{3}{*}{$\eta$=1} & $\Delta_{max}$ & 0.103 & 0.259 & 0.168 \\ \cline{3-6}
 & & $N^\prime$ & 7.72 & 8.36 & 2.81 \\ \cline{3-6}
 & & p & 1.00 & 0.56 & 1.00 \\ \cline{2-6}
 & \multirow{3}{*}{$\eta$=0.001} & $\Delta_{max}$ & 0.040 & 0.027 & 0.065 \\ \cline{3-6}
 & & $N^\prime$ & 8.61 & 8.39 & 7.67 \\ \cline{3-6}
 & & p & 1.00 & 1.00 & 1.00 \\ \hline

\end{tabular}

\end{table}

We conduct similar test for two subsets distinguished by whether or
not a long-term velocity trend is recognized and find the similar
result that our current sample size is inadequate to draw a
statistically significant conclusion on whether or not the the
groups are from the same parent distribution. Again, we see a hint
of a difference between cumulative distribution functions of two
subsets (Fig. \ref{fig:cdfe} right) although it is not statistically
significant. There are more systems with low-eccentricity for
subsets showing no sign of external perturbation. The maximum
difference between the cumulative distribution functions
$\Delta_{max}$ is 0.123, $N^\prime$ is 8.37 and K-S statistic is
0.999. In that case, we need an effective sample size of 119 in order to make a
statistically significant conclusion ($p$=0.05). In other cases, larger sample size is required since $\Delta_{max}$ is less.

We have shown that any difference in eccentricity distribution
depending on expected time scale for tidal circularization or the
presence of additional bodies capable of exciting inner planet's
eccentricity is not statistically significant, although this may be
a consequence of small effective sample size. The data are also
consistent with the argument that both factors play roles in
affecting the eccentricity distribution.

\section{Eccentricity Distribution}

We seek an analytical function that is able to approximate the
observed eccentricity distribution for short period single planetary
systems in the framework of Bayesian inference. For this purpose, we
first exclude systems showing long-term RV trends to reduce the
effect of perturbation on the estimated eccentricity distribution.
We also assume that the distribution of $\tau_{age}/\tau_{circ}$ in
our sample is representative of short-period single-planet systems.

Using the posterior samples of eccentricity from standard MCMC
analysis, we obtain an observed eccentricity probability density
function (pdf) $f(e)$ by summing the posterior
distributions together. While not statistically rigorous, this
provides a  simple summary of our results. Logarithmic
binning is adopted because the shape of $f(e)$ at low eccentricity
is of particular interest. The uncertainty $\sigma(e)$ for each bin
is set by assuming a Poisson distribution. Then, we use a
brute-force Bayesian analysis to find the most probable values of
parameters for the candidate eccentricity pdf $f'(e)$ that
approximates the observed eccentricity pdf. In the observed
eccentricity pdf, there is a pile-up in small eccentricities near
zero and a scatter of nonzero eccentricity less than 0.8. Therefore,
we use a mixture of two distributions: an exponential pdf
$f_{expo}(e,\lambda)=(1/\lambda)\cdot\exp(-e/\lambda)$ to represent the
pile-up of small eccentricity near zero and either a uniform
distribution or a Rayleigh distribution~\citep{Juric2008} to
represent the population with sizable eccentricities. We assume a
uniform distribution for parameters in prior $\Sigma(\vec\theta)$,
where $\vec{\theta}$ is vector containing the parameters for
$f'(e)$. Our results are not sensitive to the choice of priors.
We adopt Poisson likelihood for each bin in the form of $f_{Poisson}(n;\nu)=(\nu^n\cdot\exp(-\nu))/n!$, i.e.,
$L(e_i|\vec\theta)=f_{Poisson}\big(f^\prime(e_i|\vec\theta)\cdot
N;f(e_i)\cdot N\big)$, 
where $N$ is total number of posterior samples. The value of $f^\prime(e_i|\vec\theta)\cdot
N$ is rounded if it is not an integer. The posterior distribution of $\vec\theta$ is calculated as
$p(\vec\theta|\vec
e)=\big[\displaystyle\prod_i^M\Sigma(\vec\theta)L(e_{i}|\vec\theta)\big]
^\frac{1}{M}\big/\int\big[\displaystyle\prod_i^M\Sigma(\vec\theta)L(e_{i}|
\vec\theta)\big]^\frac{1}{M}d\vec\theta$ , where $M$ is the number
of bins. We have explored a range of bin size from 10 to 40 bins in
logarithmal space and conclude that the results of Bayesian analysis
do not change significantly with choice of bin size. Since the
plausible values of parameters for $f'(e)$ are limited, we do not
have to explore a large parameter space. Therefore, a brute-force
Bayesian analysis is practical.

\begin{figure}
\begin{center}
\includegraphics[width=8cm]{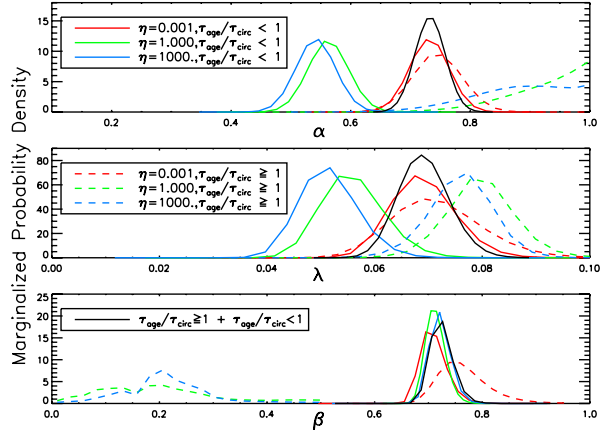}
\caption{Marginalized probability density functions of parameters
for analytical eccentricity distribution with a mixture of
exponential and uniform pdf. Uncircularized systems (group 2) are
represented by colored solid lines while circularized systems
(group 1) are represented by colored dashed lines, red:
$\eta=0.001$; green: $\eta=1$; blue: $\eta=1000$. Different colors
indicate different $\eta$ values. Solid black lines are for union of
group 1 and 2.  \label{fig:Unif_Marginalized_pdf}}
\end{center}
\end{figure}

We apply Bayesian analysis to three different planet populations for
different $\eta$ values ($\eta$=0.001,1,1000) assuming
$Q^\prime_\ast=10^7$ and $Q^\prime_P=10^7$: 1) systems without a
long-term RV slope and $\tau_{age}/\tau_{circ}\ge1$; 2) systems
without long-term RV slope and $\tau_{age}/\tau_{circ}<1$; 3) the
union of 1) and 2). The results of Bayesian analysis are presented
in Table \ref{tab:Bayesian}.
\begin{figure}
\begin{center}
\includegraphics[width=8cm]{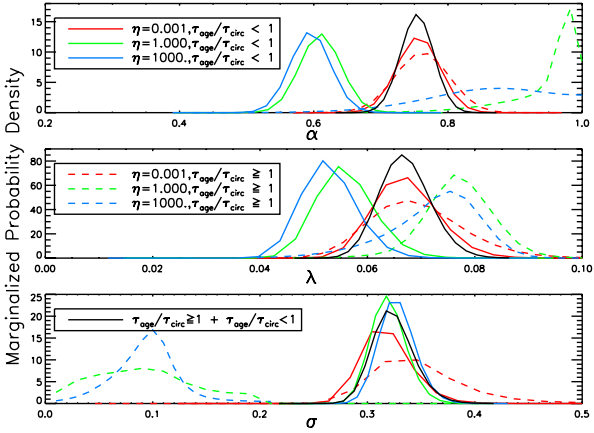}
\caption{Marginalized probability density functions of parameters
for analytical eccentricity distribution with a mixture of
exponential and Rayleigh pdfs. Uncircularized systems (group 2) are
represented by colored solid lines while circularized systems
(group 1) are represented by colored dashed lines. Different colors
indicate different $\eta$ values, red: $\eta=0.001$; green:
$\eta=1$; blue: $\eta=1000$. Solid black lines are for union of
group 1 and 2. \label{fig:Rayl_Marginalized_pdf}}
\end{center}
\end{figure}
Fig \ref{fig:Rayl_Marginalized_pdf} shows marginalized
probability density of different parameters for analytical
eccentricity distribution with a mixture of exponential and Rayleigh
distributions ($f'(e)=\alpha\cdot f_{expo}(e,\lambda)+(1-\alpha)\cdot f_{rayl}(e,\sigma)$). $\alpha$ is the fraction of exponential distribution component and $f_{rayl}(e,\sigma)$ represents Rayleigh distribution with the form of $f_{rayl}(e,\sigma)=(e/\sigma^2)\cdot\exp{(-e^2/2\sigma^2)}$. Group 1 (potentially circularized system) is represented by
dashed lines of different colors indicating different values of
$\eta$ while group 2 (systems unlikely to have been circularized) is represented by
colored solid lines. Group 1 and group 2 are well mixed if
$\eta$=0.001, i.e., a loose constraint is applied on the boundary of
$\tau_{age}/\tau_{circ}=1$. When adopting $\eta$=0.001, the marginalized posterior pdfs for both groups 1 and 2 approach the marginalized pdf for group 3 (solid black
lines; the union of group 1 and 2, including all planets
without a velocity slope). When we adopt larger $\eta$ values
($\eta$=1 or 1000), $\alpha$, the fraction of the exponential component of the pdf is greater for group
1 (Fig \ref{fig:Rayl_Marginalized_pdf} top: blue and green dashed lines) than group 2 (Fig \ref{fig:Rayl_Marginalized_pdf} top: blue and green solid lines).  This is consistent with the hypothesis that significant tidal
circularization affected group 1.  For group 1, the
fraction of the exponential component of the pdf is consistent with unity, implying that
the eccentricity distribution for planets from group 1 can be described by an exponential pdf. In comparison, there
is a substantial fraction of Rayleigh component (40\%) for the
analytical function describing eccentricity distribution for
uncircularized systems (i.e., group 2). Similar conclusions are also drawn for the
analytical eccentricity distribution with a mixture of exponential
and uniform distribution (Fig \ref{fig:Unif_Marginalized_pdf}) in the form of $f'(e)=\alpha\cdot f_{expo}(e,\lambda)+(1-\alpha)\cdot f_{unif}(e,\beta)$, where $f_{unif}(e,\beta)$ is uniform distribution with lower boundary of 0 and upper boundary of $\beta$.
\begin{figure}
\begin{center}
\includegraphics[width=8cm]{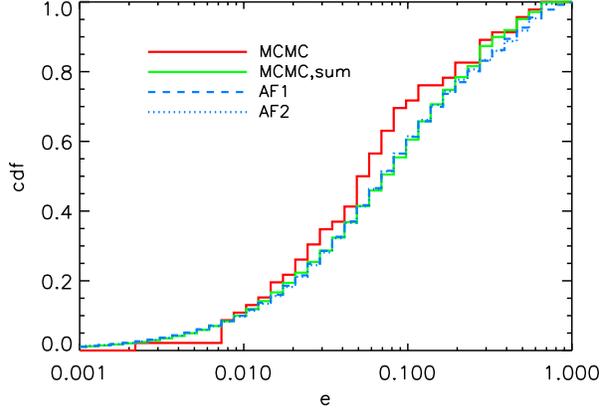}
\caption{Cumulative distributions functions (cdf) of eccentricities from different methods. MCMC: most probable eccentricities, $e_{MCMC}$, adopted from MCMC method (Appendix \ref{app:comp}); MCMC, sum: eccentricities by summing up posterior distribution samples of each system; AF1: cdf of the analytical function with the most probable parameters, $f'(e)=\alpha\cdot
f_{expo}(e,\lambda)+(1-\alpha)\cdot f_{rayl}(e,\sigma)$; AF2: cdf of the analytical function with the most probable parameters, $f'(e)=\alpha\cdot
f_{expo}(e,\lambda)+(1-\alpha)\cdot f_{unif}(e,\beta)$. \label{fig:cdf_af}}
\end{center}
\end{figure}
Fig. \ref{fig:cdf_af} shows the cumulative distribution of the eccentricities sample
based on summing the posterior eccentricity samples of each system
(green solid line) and the CDFs of the two analytic functions (blue
dotted and dashed) with the parameters that maximize the posterior
probability (see Table \ref{tab:Bayesian}, subset 3).  For comparison, the cumulative
distribution of $e_{MCMC}$ from standard MCMC analysis (Appendix \ref{app:comp}) is
shown in red. The difference between the red and the green line cannot be distinguished at 0.05 significant level in a K-S test ($N^\prime$=46).

\begin{table*}

\caption{Bayesian analysis results. Group 1: systems without
long-term RV slope and $\tau_{age}/\tau_{circ}\ge1$; group 2:
systems without long-term RV slope and $\tau_{age}/\tau_{circ}<1$;
group 3: union of 1 and 2. Numbers in bracket are uncertainties of
the last two digits. The $fraction$ column reports the percentage of
trials in which eccentricity testing samples generated by analytical
function can not be differentiated from the observed eccentricity
sample at 0.05 confidence level. \label{tab:Bayesian}}
\begin{tabular}{|c|c|c|c|c|c|c|c|c|c|}
\hline \multirow{2}{*}{$\eta$} & \multirow{2}{*}{subset} &
\multicolumn{4}{|c|}{$f'(e)=\alpha\cdot
f_{expo}(e,\lambda)+(1-\alpha)\cdot f_{rayl}(e,\sigma)$} &
\multicolumn{4}{|c|}{$f'(e)=\alpha\cdot
f_{expo}(e,\lambda)+(1-\alpha)\cdot f_{unif}(e,\beta)$} \\
\cline{3-10}  & & $\lambda[\times10^{-2}]$ & $\alpha[\times10^{-1}]$ & $\sigma[\times10^{-1}]$ & fraction & $\lambda[\times10^{-2}]$ & $\alpha[\times10^{-1}]$ & $\beta[\times10^{-1}]$ & fraction  \\
\hline \multirow{2}{*}{0.001} & 1 & 6.83(84) & 7.61(40) & 3.38(36) & 100.0\% & 7.08(82) & 7.35(38) & 7.56(41) & 100.0\% \\
\cline{2-10} & 2 & 6.67(59) & 7.56(32) & 3.15(22) & 100.0\% & 6.86(59) & 7.32(33) & 7.06(23) & 100.0\% \\
\hline \multirow{2}{*}{1.000} & 1 & 7.71(58) & 9.78(20) & 0.83(54) & 99.6\% & 7.96(59) & 10.0(12) & 1.8(11) & 99.9\% \\
\cline{2-10} & 2 & 5.55(53) & 6.10(31) & 3.19(16) & 100.0\% & 5.55(57) & 5.66(34) & 7.12(18) & 100.0\% \\
\hline \multirow{2}{*}{1000.} & 1 & 7.41(72) & 8.9(12) & 0.95(24) & 99.9\% & 7.62(56) & 9.2(11) & 2.10(57) & 100.0\% \\
\cline{2-10} & 2 & 5.24(49) & 5.95(30) & 3.27(16) & 100.0\% & 5.11(53) & 5.44(33) & 7.20(19) & 100.0\% \\
\hline & 3 & 6.67(47) & 7.54(25) & 3.22(19) & 100.0\% & 6.89(47) & 7.33(26) & 7.22(20) & 100.0\% \\
\hline

\end{tabular}

\end{table*}

In order to check whether the analytical function with the most
probable parameters is an acceptable approximation to the
eccentricity distribution for short-period single-planet systems, we
generate test samples from the analytical functions and compare the
resulting eccentricity samples to observations. For each test sample
there are $N^\prime$ eccentricities following the distribution of
the analytical function $f'(e|\hat{\theta})$, where $N^\prime$ is the effective sample size. Each eccentricity is
perturbed by an simulated measurement error that follows the
distribution of posterior samples for our analysis of the actual
observations. The test sample is then compared to the observed
eccentricity samples obtained by standard MCMC analysis using
two-sample K-S test. We report in Table \ref{tab:Bayesian} the
percentage of trials in which eccentricity samples generated by our
analytical function can not be differentiated from the observed
eccentricity sample at 0.05 confidence level. All the candidate
analytical functions we have tested are able to reproduce the
observed eccentricity distribution in more than 99.6\% of the trials
at 0.05 confidence level. We conclude that the best-fit analytical
function is an adequate approximation to observed eccentricity
distribution. We also compare to analytical eccentricity
distribution used in ~\citet{Shen2008} $f'(e)\sim[1/(1+e)^a-e/2^a]$
in which a=4, although it is not specifically for short period
single planetary systems. Similar to what we did in previous test,
we found that in 39.0\% of the tests, the eccentricity samples
generated by analytical functions can not be differentiated from the
observed eccentricity sample at 0.05 confidence level.

From the results of Bayesian analysis, there is a clear difference
in $\alpha$, the fraction of exponential distribution, between
group 1 and 2, suggesting the role played by tidal
circularization. Group 1 with $\tau_{age}/\tau_{circ}\ge1$
shows more planets with near-zero eccentricities ($\alpha\sim75\%$)
as compared to group 2 ($\alpha\sim55\%$) with
$\tau_{age}/\tau_{circ}<1$ (Table \ref{tab:Bayesian}). Since the eccentricity samples tested
were perturbed by measurement errors, the analytical function we
found can be interpreted as an approximation of the underlying
eccentricity distribution of short-period single-planet systems.
However, the actual parameter values and uncertainties in the
analytical function are dependent upon the quality of observation
and the number of systems in the sample of short-period single
planets. As the measurement precision and the sample size improve,
we will be able to better constrain the values of parameters in the
analytical function which approximates the underlying eccentricity
distribution.

\section{Discussion}

\begin{figure}
\begin{center}
\includegraphics[width=8cm]{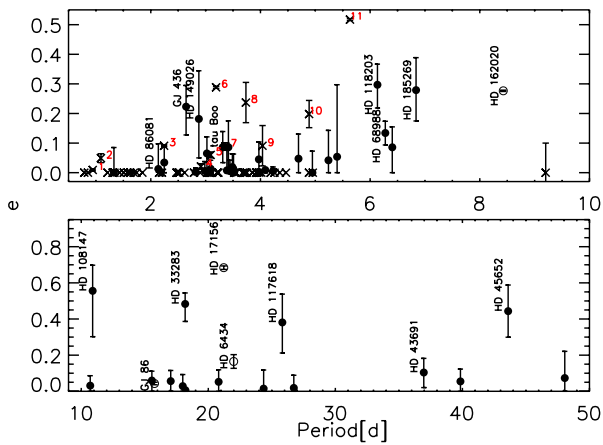}
\caption{Distribution of short-period single-planet systems in our
sample in period-eccentricity space (top: p $\le$ 10 day; bottom: p
$>$ 10 day). Systems that are not consistent with zero eccentricity
according to $\Gamma$ analysis (filled circles) or MCMC analysis
(open circles) are marked with corresponding names. We also include
transiting planets (marked as cross) for comparison. Transiting
systems not consistent with zero eccentricity are marked with
numbers: 1, WASP-18; 2, WASP-12; 3, WASP-14; 4, HAT-P-13; 5,
WASP-10; 6, XO-3; 7, WASP-6; 8,WASP-17; 9, CoRoT-5; 10, HAT-P-11;
11, HAT-P-2. \label{fig:pe}}
\end{center}
\end{figure}

The median eccentricity of short-period single-planet systems in our
sample is 0.088. When compared to median eccentricity for all the
detected exoplanets 0.15, it suggests that the population may be
affected by tidal circularization. We use eccentricity estimated by
$\Gamma$ analysis in the following discussion since it is a less biased method for accessing small eccentricity. Fig.
\ref{fig:pe} shows the period-eccentricity correlation. We would
expect planets with sufficient short period to be tidally
circularized. While this is generally true, there are 3 (17.6\%)
planets with $P<4d$ and non-circular orbits: GJ 436, HD 149026 and
$\tau$ Boo. For GJ 436, it is suspected that an outer companion may
be pumping the eccentricity ~\citep{Maness2007,Ribas2008}.
Similarly, observations of $\tau$ Boo b are inconsistent with
circular orbit, but might be explained by the perturbation of an
unseen companion indicated by a long-term RV linear
trend~\citep{Butler2006}. Secondary eclipse timing indicates that
the eccentricity of HD 149026 is quite small but inconsistent with
zero~\citep{Knutson2009}. Considering planets with orbital periods
up to 10 days, there are 4 additional systems that are not
circularized. HD 118203, HD 68988 and HD 185269 might be due to
perturbations by additional bodies in the system~\citep{Butler2006}
while the non-zero eccentricity of HD 162020 b may be attributed to
a different formation mechanism~\citep{Udry2002}. With period longer
than 10 days, there are 8 (44.4\%) planets with
$\tau_{age}/\tau_{circ}>1$ which have non-zero eccentricity and no
detected long-term linear trend. In comparison, there are 8 (28.6\%)
eccentric planets with orbital periods less than 10 days. The
increasing fraction of recognizably eccentric orbits as period
increases is suggestive of decreasing tidal circularization effect, but large
sample of planets is required to draw firm conclusion. The discovery
of over 705 planets candidates by the Kepler mission presents an
excellent opportunity to analyze the eccentricity distribution of
short-period planets~\citep{Ford2008b,Borucki2010}. We briefly
consider transiting planets. We note that 11 of 58 (19.0\%) transiting systems as of June 2010
are not consistent with zero eccentricity, and the orbital periods
for all transiting planets but 3 (i.e. CoRoT-9b, HAT-P-13c HD
80606b) are less than 10 days. We infer that the
tidal circularization process might be effective for isolated
planets with orbital period of less than 10 days. Alternatively, the
planet formation and migration processes for short-period giant
planet may naturally lead to a significant fraction of nearly
circular orbits, even before tidal effect takes place.

It is worth noting that HD 17156 b~\citep{Fischer2007}
has one of the most eccentric orbits among short-period planet systems
in spite the fact that $\tau_{age}/\tau_{circ}$ could be well over
10 (Fig \ref{fig:eCircAge}, red). However, ~\citet{Barbieri2009}
found no indications of additional companions based on observations
of directing imaging, RV and astrometry measurement.
~\citet{Anglada2010} investigated the possibility that a 2:1
resonant orbit can be hidden by an eccentric orbital solution. It is
interesting to explore such possibility on this particular system to
solve the discrepancy of high eccentricity and $\tau_{age}$ to
$\tau_{circ}$ ratio. Another possibility is that the system is in
the process of circularization that began well after the star and
planet formed (e.g., due to planets scattering). However, we are
cautious in drawing conclusions since $\tau_{age}/\tau_{circ}$
could be less than unity (Fig \ref{fig:eCircAge}, green and blue).

\section{Conclusion}

We apply standard MCMC analysis for 50 short-period single-planet
systems and construct a catalog of orbital solutions for these
planetary systems (Appendix \ref{app:cat}). We find general
agreement between MCMC analysis and previous references with the
primary exception being cases where eccentricity was held fixed in
previous analysis. We develop a new method to test the significance
of non-circular orbits ($\Gamma$ analysis), which is better suited
to performing population analysis. We find the eccentricity
estimations from $\Gamma$ analysis are consistent with results from
both standard MCMC analysis and previous references.

Our results suggest that both tidal interactions and external
perturbations may play roles in shaping the eccentricity
distribution of short-period single-planet systems but large sample
sizes are needed to provide sufficient sensitivity to make these
trends statistically significant. We identify two analytical
functions that approximate the underlying eccentricity distribution:
1) mixture of an exponential distribution and a uniform distribution
and 2) a mixture of an exponential distribution and a Rayleigh
distribution. We use Bayesian analysis to find the most probable
values of parameters for the analytical functions given the observed
eccentricities (Table \ref{tab:Bayesian}). The analytical functions
can be interpreted as the underlying distribution of eccentricities
for short-period single-planet systems. Thus, the analytical
functions can be used in the future theoretical works or as priors for
eccentricity distribution.

\section*{Acknowledgments}
We thank Matthew Payne, Margaret Pan and Nadia Zakamska for useful
discussions. This research was partially supported by NASA Origins
of Solar Systems grant NNX09AB35G. This material is based upon work
supported by the National Science Foundation under Grant No.
0707203.

\bibliographystyle{mn2e}
\bibliography{mybib_JW_EF}{}

\onecolumn

\begin{landscape}
\appendix
\section{}
\subsection{Comparison of Eccentricities Calculated From Different
Methods} \label{app:comp} \scriptsize

\begin{longtable}{lcccccccc}
\hline
Name & \multicolumn{2}{c}{Ref.} & \multicolumn{3}{c}{MCMC} & \multicolumn{3}{c}{$\Gamma$} \\
\hline
 & $e_{ref}$ & $\delta e_{ref}$ & $e_{MCMC}$ & $e_{lower}$ & $e_{upper}$ & $e_{\Gamma}$ & $e_{lower}$ & $e_{upper}$ \\
\hline
HD 41004 B & 0.081 & 0.012 & 0.058 & 0.000 & 0.109 & 0.000 & 0.000 & 0.085 \\
HD 86081 & 0.008 & 0.004 & 0.058 & 0.008 & 0.166 & 0.013 & 0.001 & 0.098 \\
HD 189733 & 0.000 & 0.000 & ... & ... & ... & ... & ... & ... \\
HD 212301 & 0.000 & ... & 0.015 & 0.000 & 0.051 & 0.034 & 0.000 & 0.091  \\
GJ 436 & 0.159 & 0.052 & 0.191 & 0.146 & 0.248 & 0.223 & 0.128 & 0.295  \\
HD 63454 & 0.000 & ... & 0.018 & 0.000 & 0.038 & 0.003 & 0.000 & 0.017  \\
HD 149026 & 0.000 & ... & 0.192 & 0.118 & 0.270 & 0.182 & 0.050 & 0.344  \\
HD 83443 & 0.012 & 0.023 & 0.007 & 0.000 & 0.020 & 0.006 & 0.000 & 0.037  \\
HD 46375 & 0.063 & 0.026 & 0.052 & 0.030 & 0.084 & 0.065 & 0.000 & 0.121  \\
HD 179949 & 0.022 & 0.015 & 0.014 & 0.000 & 0.024 & 0.013 & 0.000 & 0.042  \\
$\tau$ Boo & 0.023 & 0.015 & 0.079 & 0.054 & 0.117 & 0.086 & 0.034 & 0.139  \\
HD 330075 & 0.000 & ... & 0.019 & 0.000 & 0.087 & 0.008 & 0.000 & 0.095  \\
HD 88133 & 0.133 & 0.072 & 0.076 & 0.000 & 0.127 & 0.086 & 0.000 & 0.175  \\
HD 2638 & 0.000 & ... & 0.041 & 0.000 & 0.076 & 0.005 & 0.000 & 0.042  \\
BD -10 3166 & 0.019 & 0.023 & 0.010 & 0.000 & 0.030 & 0.019 & 0.000 & 0.064  \\
HD 75289 & 0.034 & 0.029 & 0.021 & 0.000 & 0.043 & 0.000 & 0.000 & 0.063  \\
HD 209458 & 0.000 & ... & 0.008 & 0.000 & 0.016 & 0.005 & 0.000 & 0.018  \\
HD 219828\footnotemark[1] & 0.000 & ... & ... & ... & ... & ... & ... & ... \\
HD 76700 & 0.095 & 0.075 & 0.062 & 0.003 & 0.104 & 0.045 & 0.000 & 0.104  \\
HD 149143 & 0.000 & ... & 0.012 & 0.000 & 0.022 & 0.009 & 0.000 & 0.014  \\
HD 102195 & 0.000 & ... & ... & ... & ... & ... & ... & ... \\
51 Peg & 0.013 & 0.012 & 0.007 & 0.000 & 0.014 & 0.006 & 0.000 & 0.020  \\
GJ 674\footnotemark[2] & 0.100 & 0.020 & 0.070 & 0.000 & 0.147 & 0.047 & 0.000 & 0.131  \\
HD 49674 & 0.087 & 0.095 & 0.050 & 0.000 & 0.119 & 0.000 & 0.000 & 0.073  \\
HD 109749 & 0.000 & ... & 0.045 & 0.000 & 0.070 & 0.042 & 0.000 & 0.143  \\
HD 7924 & 0.170 & 0.160 & 0.119 & 0.022 & 0.224 & 0.054 & 0.000 & 0.297  \\
HD 118203 & 0.309 & 0.014 & 0.293 & 0.264 & 0.328 & 0.297 & 0.218 & 0.367  \\
HD 68988 & 0.125 & 0.009 & 0.118 & 0.096 & 0.143 & 0.134 & 0.094 & 0.174  \\
HD 168746 & 0.107 & 0.080 & 0.079 & 0.025 & 0.139 & 0.086 & 0.000 & 0.155  \\
HD 185269\footnotemark[3] & 0.276 & 0.037 & 0.276 & 0.242 & 0.314 & 0.279 & 0.175 & 0.389  \\
HD 162020 & 0.277 & 0.002 & 0.277 & 0.274 & 0.279 & ... & ... & ...  \\
GJ 176\footnotemark[4] & 0.000 & ... & ... & ... & ... & ... & ... & ...  \\
HD 130322 & 0.011 & 0.020 & 0.007 & 0.000 & 0.052 & 0.031 & 0.000 & 0.086  \\
HD 108147 & 0.530 & 0.120 & 0.526 & 0.429 & 0.624 & 0.556 & 0.302 & 0.698  \\
HD 4308 & 0.000 & 0.010 & 0.068 & 0.000 & 0.123 & 0.060 & 0.000 & 0.111  \\
GJ 86 & 0.042 & 0.007 & 0.042 & 0.034 & 0.051 & ... & ... & ...  \\
HD 99492 & 0.050 & 0.120 & 0.036 & 0.000 & 0.125 & 0.056 & 0.000 & 0.115  \\
HD 27894 & 0.049 & 0.008 & 0.024 & 0.000 & 0.045 & 0.029 & 0.000 & 0.093  \\
HD 33283 & 0.480 & 0.050 & 0.458 & 0.401 & 0.523 & 0.483 & 0.386 & 0.544  \\
HD 195019 & 0.014 & 0.004 & 0.002 & 0.000 & 0.007 & 0.006 & 0.000 & 0.014  \\
HD 102117 & 0.121 & 0.082 & 0.068 & 0.038 & 0.121 & 0.052 & 0.000 & 0.119  \\
HD 17156\footnotemark[5] & 0.684 & 0.013 & 0.683 & 0.672 & 0.691 & ... & ... & ...  \\
HD 6434 & 0.170 & 0.030 & 0.159 & 0.124 & 0.202 & ... & ... & ...  \\
HD 192263 & 0.055 & 0.039 & 0.026 & 0.000 & 0.055 & 0.015 & 0.000 & 0.118  \\
HD 117618 & 0.420 & 0.170 & 0.352 & 0.233 & 0.511 & 0.381 & 0.212 & 0.538  \\
HD 224693 & 0.050 & 0.030 & 0.031 & 0.000 & 0.055 & 0.019 & 0.000 & 0.089  \\
HD 43691\footnotemark[6] & 0.140 & 0.020 & 0.090 & 0.055 & 0.133 & 0.104 & 0.021 & 0.182  \\
$\rho$ Crb & 0.057 & 0.028 & 0.048 & 0.000 & 0.069 & 0.055 & 0.000 & 0.123  \\
HD 45652\footnotemark[7] & 0.380 & 0.060 & 0.434 & 0.371 & 0.496 & 0.443 & 0.299 & 0.588  \\
HD 107148 & 0.050 & 0.170 & 0.028 & 0.000 & 0.141 & 0.073 & 0.000 & 0.221  \\
\end{longtable}
{References: 1~\citet{Melo2007}; 2~\citet{Bonfils2007};
3~\citet{Johnson2006}; 4~\citet{Forveille2009};
5~\citet{Barbieri2009}; 6~\citet{daSilva2007}; 7~\citet{Santos2008};
$e_{ref}$ and $\delta e_{ref}$ are from ~\citet{Butler2006} if
otherwise noted.}
\subsection{Catalog of Short-Period Single-Planet Systems}
\label{app:cat} \scriptsize
\begin{longtable}{lccccccccccccccccc}
  \hline
Name & $P$ & $K$ & $e$ & $\omega$ & $M0$ & $\tau$ & trend & Jitter & $N_{obs}$ & $M_*$ & $R_*$ & $M_P$ & $R_P$ & $\tau_{circ}$ & $\tau_{age}$ & RV \\
 & (day) & ($\rm{m}\cdot\rm{s}^{-1}$) & & (deg) & (deg) & (day) & ($\rm{m}\cdot\rm{s}^{-1}\cdot\rm{d}^{-1}$) & ($\rm{m}\cdot\rm{s}^{-1}$) & & ($M_\bigodot$) & ($R_\bigodot$) & ($M_J$) & ($R_J$) & (Gyr) & (Gyr) & ref. \\
 \hline
HD 41004B  &   1.32363$\pm$8.91594e-05  &   4599.21$\pm$337.57  &   0.0580$^{+0.0511}_{-0.0580}$  &   149.0$\pm$72.2  &   119.5$\pm$72.1  &   2452532.699  &   ...$\pm$...  &   2761.45$\pm$168.90  &   149   &  0.40  &  0.40  &  18.40  &  1.06  &  0.26  &  6.32 & 7  \\
HD 86081  &   1.99809$\pm$0.00677822  &   189.65$\pm$12.09  &   0.0575$^{+0.1080}_{-0.0501}$  &   -27.4$\pm$71.5  &   64.8$\pm$74.5  &   2453753.2  &   ...$\pm$...  &   32.13$\pm$5.74  &   26   &  1.21  &  1.22  &  1.50  &  1.08  &  0.56  &  6.21 & 8  \\
HD 189733\footnotemark[1] &  2.2186$\pm$0.0005 &  205$\pm$6 &  0$\pm$0.0002 &  90 &  270 &  2454037.612 &  ...$\pm$... &  15 &  86 & ... & ... & ... & ... & ... & ... & 9 \\
HD 212301  &   2.24571$\pm$0.000147699  &   57.26$\pm$3.01  &   0.0147$^{+0.0365}_{-0.0147}$  &   172.8$\pm$85.6  &   56.0$\pm$85.6  &   2453388.9  &   ...$\pm$...  &   8.53$\pm$1.78  &   23   &  1.05  &  1.19  &  0.40  &  1.07  &  0.28  &  5.90 & 11  \\
GJ 436  &   2.64394$\pm$9.85041e-05  &   18.07$\pm$1.03  &   0.1912$^{+0.0571}_{-0.0449}$  &   -5.6$\pm$15.3  &   -66.6$\pm$14.5  &   2452992.1  &   0.0037$\pm$0.001\footnotemark[5]  &   0.41$\pm$1.46  &   55   &  0.41  &  0.46  &  0.07  &  0.38  &  5.57  &  6.00 & 6  \\
HD 63454  &   2.81747$\pm$0.000382247  &   63.19$\pm$1.82  &   0.0177$^{+0.0203}_{-0.0177}$  &   -122.9$\pm$76.2  &   25.7$\pm$76.3  &   2453238.057  &   ...$\pm$...  &   5.70$\pm$1.22  &   26   &  0.80  &  0.78  &  0.39  &  1.06  &  0.67  &  1.00 & 12  \\
HD 149026  &   2.87807$\pm$0.00146571  &   54.63$\pm$11.90  &   0.1918$^{+0.0777}_{-0.0743}$  &   114.2$\pm$25.9  &   10.9$\pm$22.5  &   2453545.35  &   ...$\pm$...  &   2.41$\pm$3.26  &   17   &  1.30  &  1.50  &  0.36  &  0.65  &  3.46  &  2.00 & 6  \\
HD 83443  &   2.98572$\pm$5.30373e-05  &   56.00$\pm$1.05  &   0.0070$^{+0.0135}_{-0.0070}$  &   117.3$\pm$82.2  &   134.0$\pm$82.0  &   2452248.9  &   ...$\pm$...  &   3.12$\pm$1.67  &   51   &  1.00  &  1.02  &  0.40  &  1.04  &  1.10  &  11.68 & 6  \\
HD 46375  &   3.02358$\pm$6.44902e-05  &   33.67$\pm$0.81  &   0.0524$^{+0.0320}_{-0.0229}$  &   113.7$\pm$34.3  &   -53.2$\pm$34.3  &   2451920.7  &   ...$\pm$...  &   3.28$\pm$0.60  &   50   &  0.92  &  0.94  &  0.23  &  1.02  &  0.67  &  11.88 & 6  \\
HD 179949  &   3.0925$\pm$3.30046e-05  &   112.62$\pm$1.77  &   0.0104$^{+0.0099}_{-0.0104}$  &   -170.7$\pm$63.4  &   42.4$\pm$63.3  &   2452419.1  &   ...$\pm$...  &   9.44$\pm$1.06  &   88   &  1.21  &  1.22  &  0.92  &  1.05  &  2.90  &  2.56 & 6  \\
$\tau$ Boo  &   3.31249$\pm$3.12595e-05  &   469.59$\pm$14.86  &   0.0787$^{+0.0382}_{-0.0246}$  &   -141.6$\pm$25.0  &   24.4$\pm$24.9  &   2450529.2  &   -0.051$\pm$0.003\footnotemark[6]  &   94.30$\pm$8.13  &   98   &  1.35  &  1.33  &  4.13  &  1.06  &  4.12  &  1.64 & 6  \\
HD 88133  &   3.41566$\pm$0.000841134  &   34.13$\pm$3.57  &   0.0761$^{+0.0508}_{-0.0761}$  &   -2.8$\pm$60.4  &   -39.5$\pm$60.1  &   2453180.0  &   ...$\pm$...  &   5.67$\pm$1.61  &   21   &  1.20  &  1.93  &  0.30  &  1.00  &  1.49  &  9.56 & 6  \\
HD 2638  &   3.43752$\pm$0.00823876  &   66.26$\pm$2.83  &   0.0407$^{+0.0351}_{-0.0407}$  &   126.9$\pm$78.6  &   -123.4$\pm$77.7  &   2453323.282  &   ...$\pm$...  &   5.48$\pm$4.87  &   28   &  0.93  &  1.01  &  0.48  &  1.04  &  2.38  &  3.00 & 12  \\
BD -10 3166  &   3.4878$\pm$0.000104858  &   60.53$\pm$1.44  &   0.0104$^{+0.0192}_{-0.0104}$  &   -14.6$\pm$83.9  &   36.6$\pm$83.7  &   2451844.7  &   0.005$\pm$0.002\footnotemark[6]  &   4.00$\pm$1.74  &   31   &  1.01  &  0.84  &  0.46  &  1.03  &  2.70  &  1.84 & 6  \\
HD 75289  &   3.50928$\pm$7.2946e-05  &   54.84$\pm$1.87  &   0.0211$^{+0.0217}_{-0.0211}$  &   136.3$\pm$73.9  &   -162.2$\pm$74.1  &   2452593.9  &   ...$\pm$...  &   4.73$\pm$1.73  &   30   &  1.21  &  1.28  &  0.47  &  1.03  &  3.03  &  3.28 & 6  \\
HD 209458  &   3.52472$\pm$2.81699e-05  &   84.30$\pm$0.88  &   0.0082$^{+0.0078}_{-0.0082}$  &   43.8$\pm$68.4  &   92.5$\pm$68.5  &   2452499.3  &   ...$\pm$...  &   3.27$\pm$0.86  &   64   &  1.14  &  1.14  &  0.69  &  1.05  &  4.07  &  2.44 & 6  \\
HD 330075  &   3.6413$\pm$0.00187111  &   97.84$\pm$8.79  &   0.0187$^{+0.0684}_{-0.0187}$  &   38.2$\pm$91.6  &   35.2$\pm$91.7  &   2452968.399  &   ...$\pm$...  &   24.59$\pm$5.00  &   21   &  0.70  &  0.90  &  0.62  &  1.06  &  1.97  &  6.21 & 13  \\
HD 219828\footnotemark[2] &  3.833$\pm$0.0013 &  7$\pm$0.5 &  0 &  0 &  0 &  2453898.63 &  ...$\pm$... &  1.7 &  27 & ... & ... & ... & ... & ... & ... & 2  \\
HD 76700  &   3.97101$\pm$0.000203194  &   27.24$\pm$1.31  &   0.0616$^{+0.0426}_{-0.0587}$  &   12.3$\pm$54.3  &   45.6$\pm$53.8  &   2452655.1  &   ...$\pm$...  &   1.35$\pm$4.21  &   35   &  1.13  &  1.34  &  0.23  &  0.99  &  2.80  &  9.84 & 6  \\
HD 149143  &   4.07206$\pm$0.000320041  &   149.28$\pm$1.65  &   0.0123$^{+0.0093}_{-0.0115}$  &   -155.2$\pm$55.9  &   -150.9$\pm$55.9  &   2453413.1  &   0.027$\pm$0.005\footnotemark[6]  &   0.48$\pm$1.96  &   17   &  1.20  &  1.61  &  1.33  &  1.05  &  7.01  &  7.60 & 14  \\
HD 102195\footnotemark[3] &  4.1138$\pm$0.000557 &  63$\pm$2 &  0 &  0 &  0 &  2453895.96 &  ...$\pm$... &  6.1 &  59 & ... & ... & ... & ... & ... & ... & 3  \\
51 PEG  &   4.2308$\pm$3.72905e-05  &   55.65$\pm$0.53  &   0.0069$^{+0.0066}_{-0.0069}$  &   54.1$\pm$72.3  &   85.1$\pm$72.3  &   2450404.4  &   -0.0045$\pm$0.0004\footnotemark[6]  &   0.27$\pm$0.91  &   256   &  1.09  &  1.18  &  0.47  &  1.03  &  6.61  &  6.76 & 6  \\
GJ 674  &   4.6944$\pm$0.00182591  &   9.46$\pm$1.09  &   0.0700$^{+0.0766}_{-0.0700}$  &   0.5$\pm$71.2  &   77.6$\pm$71.1  &   2453823.784  &   ...$\pm$...  &   3.55$\pm$0.55  &   32   &  0.35  &  0.46  &  0.04  &  1.13  &  0.21  &  0.55 & 15  \\
HD 49674  &   4.94739$\pm$0.000974925  &   11.78$\pm$1.18  &   0.0495$^{+0.0691}_{-0.0495}$  &   -96.1$\pm$79.3  &   82.4$\pm$79.3  &   2452308.9  &   ...$\pm$...  &   3.56$\pm$0.82  &   39   &  1.06  &  0.95  &  0.10  &  0.98  &  3.25  &  3.56 & 6  \\
HD 109749  &   5.23921$\pm$0.000935533  &   28.49$\pm$1.12  &   0.0451$^{+0.0250}_{-0.0451}$  &   72.3$\pm$55.8  &   -155.1$\pm$55.8  &   2453426.3  &   ...$\pm$...  &   0.31$\pm$1.10  &   20   &  1.21  &  1.28  &  0.28  &  0.99  &  12.04  &  10.30 & 14  \\
HD 7924  &   5.39785$\pm$0.00096697  &   3.74$\pm$0.44  &   0.1186$^{+0.1050}_{-0.0970}$  &   25.0$\pm$55.6  &   62.1$\pm$54.8  &   2454096.65  &   0.35$\pm$0.07  &   2.59$\pm$0.22  &   93   &  0.83  &  0.78  &  0.03  &  1.05  &  0.61  &  0.88 & 16  \\
HD 118203  &   6.13322$\pm$0.00129898  &   213.75$\pm$6.67  &   0.2943$^{+0.0342}_{-0.0298}$  &   -27.3$\pm$5.5  &   -2.9$\pm$4.8  &   2453351.2  &    0.14$\pm$0.016\footnotemark[6]  &   23.08$\pm$3.96  &   43   &  1.23  &  2.15  &  2.14  &  1.05  &  3.68  &  4.60 & 17  \\
HD 68988  &   6.27699$\pm$0.0002195  &   184.63$\pm$4.69  &   0.1187$^{+0.0246}_{-0.0216}$  &   32.4$\pm$11.2  &   61.3$\pm$10.8  &   2452441.3  &   -0.065$\pm$0.005\footnotemark[6]  &   13.30$\pm$2.34  &   28   &  1.18  &  1.14  &  1.86  &  1.05  &  80.15  &  3.40 & 6  \\
HD 168746  &   6.40398$\pm$0.000979461  &   28.41$\pm$1.38  &   0.0791$^{+0.0595}_{-0.0541}$  &   14.8$\pm$47.2  &   -147.2$\pm$47.3  &   2452510.8  &   ...$\pm$...  &   0.41$\pm$1.62  &   16   &  0.93  &  1.04  &  0.25  &  0.99  &  20.31  &  12.40 & 6     \\
HD 185269  &   6.83796$\pm$0.00119146  &   89.39$\pm$4.20  &   0.2758$^{+0.0386}_{-0.0334}$  &   173.3$\pm$6.4  &   -99.4$\pm$5.9  &   2453795.0  &   ...$\pm$...  &   7.70$\pm$2.10  &   30   &  1.28  &  1.88  &  0.94  &  1.04  &  19.34 &  4.20  & 8 \\
HD 162020  &   8.42826$\pm$7.76104e-05  &   1808.84$\pm$5.26  &   0.2765$^{+0.0023}_{-0.0025}$  &   -151.2$\pm$0.3  &   68.3$\pm$1.0  &   2451672.02  &   ...$\pm$...  &   11.02$\pm$3.43  &   46   &  0.78  &  0.74  &  15.00  &  0.98  &  148.00  &  0.76 & 19  \\
GJ 176\footnotemark[4] &  8.783$\pm$0.0054 &  4.1$\pm$0.52 &  0 &  0 &  0 &  2454399.8 &  ...$\pm$... &  2.5 &  57 & ... & ... & ... & ... & ... & ... & 4  \\
HD 130322  &   10.7086$\pm$0.00184045  &   108.10$\pm$7.31  &   0.0068$^{+0.0455}_{-0.0068}$  &   128.3$\pm$93.9  &   -10.4$\pm$93.8  &   2452430.2  &   ...$\pm$...  &   10.25$\pm$3.79  &   12   &  0.88  &  0.85  &  1.09  &  1.04  &  670.08  &  10.80 & 6  \\
HD 108147  &   10.8984$\pm$0.00316767  &   24.60$\pm$3.62  &   0.5161$^{+0.0979}_{-0.0966}$  &   -54.4$\pm$14.0  &   -68.0$\pm$9.9  &   2452407.0  &   ...$\pm$...  &   8.65$\pm$1.49  &   54   &  1.19  &  1.25  &  0.26  &  0.98  &  8.43  &  3.20 & 6  \\
HD 4308  &   15.5646$\pm$0.0213556  &   4.20$\pm$0.34  &   0.0682$^{+0.0547}_{-0.0682}$  &   -166.9$\pm$60.6 &   -13.0$\pm$59.5  &   2453338.121  &   ...$\pm$...  &   0.58$\pm$0.78  &   41   &  0.90  &  0.92  &  0.05  &  1.00  &  169.24  &  8.68 & 20 \\
GJ 86  &   15.765$\pm$0.000382114  &   376.64$\pm$2.79  &   0.0416$^{+0.0092}_{-0.0073}$  &   -93.7$\pm$12.4  &   -76.8$\pm$12.0  &   2452199.4  &   -0.260$\pm$0.0029  &   10.67$\pm$1.62  &   42   &  0.77  &  0.80  &  3.91  &  1.05  &  6701.62  &  8.48 & 6  \\
HD 99492  &   17.0495$\pm$0.00525091  &   8.39$\pm$0.98  &   0.0364$^{+0.0891}_{-0.0364}$  &   -138.6$\pm$93.6  &   -137.6$\pm$93.5  &   2452523.85  &   0.0035$\pm$0.0006\footnotemark[6]  &   3.96$\pm$0.51  &   86   &  0.86  &  0.76  &  0.11  &  0.96  &  692.89  &  1.80 & 21  \\
HD 27894  &   18.0059$\pm$0.0163248  &   56.88$\pm$1.75  &   0.0240$^{+0.0211}_{-0.0240}$  &   131.3$\pm$66.5  &   -60.0288$\pm$66.5  &   2453344.278  &   ...$\pm$...  &   4.48$\pm$1.07  &   20   &  0.75  &  0.90  &  0.62  &  1.02  &  3620.33  &  3.90 & 12  \\
HD 33283  &   18.1801$\pm$0.00830584  &   24.48$\pm$2.41  &   0.4576$^{+0.0656}_{-0.0569}$  &   156.0$\pm$9.6  &   -60.0$\pm$7.4  &   2453560.0  &   ...$\pm$...  &   0.27$\pm$0.94  &   24   &  1.24  &  1.20  &  0.33  &  0.99  &  203.12  &  3.20 & 8  \\
HD 195019  &   18.2018$\pm$0.000595221  &   269.70$\pm$1.60  &   0.0017$^{+0.0049}_{-0.0017}$  &   -127.4$\pm$47.4  &   -175.5$\pm$47.3  &   2451844.0  &   ...$\pm$...  &   10.50$\pm$1.22  &   154   &  1.07  &  1.38  &  3.69  &  1.05  &  4051.07  &  9.32 & 6  \\
HD 102117  &   20.8210$\pm$0.01006465  &   10.20$\pm$0.91  &   0.0685$^{+0.0529}_{-0.0647}$  &   137.6$\pm$72.0  &   -9.5$\pm$72.5  &   2452931.7  &   ...$\pm$...  &   0.57$\pm$1.73  &   44   &  1.11  &  1.26  &  0.17  &  0.95  &  3205.49  &  9.40 & 6  \\
HD 17156  &   21.2178$\pm$0.00371004  &   279.88$\pm$8.43  &   0.6829$^{+0.0080}_{-0.0106}$  &   121.3$\pm$1.1  &   -156.7$\pm$1.2  &   2454111.21  &   ...$\pm$...  &   3.68$\pm$0.64  &   34   &  1.24  &  1.63  &  3.20  &  1.04  &  19.10  &  8.00 & 10  \\
HD 6434  &   21.9975$\pm$0.0127604  &   34.30$\pm$1.52  &   0.1589$^{+0.0434}_{-0.0345}$  &   155.4$\pm$15.8  &   -14.2$\pm$15.5  &   2451753.933  &   ...$\pm$...  &   7.39$\pm$1.15  &   130   &  0.79  &  0.57  &  0.40  &  1.00  &  4393.14  &  6.85 & 22  \\
HD 192263  &   24.3546$\pm$0.00507675  &   51.13$\pm$2.62  &   0.0256$^{+0.0297}_{-0.0256}$  &   -147.0$\pm$78.9  &   -56.7$\pm$78.9  &   2451867.6  &   ...$\pm$...  &   6.99$\pm$1.31  &   31   &  0.81  &  0.77  &  0.64  &  1.02  &  14630.88  &  2.56 & 6  \\
HD 117618  &   25.8221$\pm$0.0155045  &   12.25$\pm$1.70  &   0.3524$^{+0.1583}_{-0.1192}$  &   -105.6$\pm$22.9  &   -123.3$\pm$21.1  &   2452838.0  &   ...$\pm$...  &   3.30$\pm$1.20  &   57   &  1.09  &  1.20  &  0.18  &  0.95  &  1435.34  &  5.68 & 6  \\
HD 224693  &   26.732$\pm$0.0245934  &   39.73$\pm$1.54  &   0.0313$^{+0.0236}_{-0.0313}$  &   11.6$\pm$68.6  &   162.5$\pm$68.9  &   2453607.2  &   ...$\pm$...  &   1.00$\pm$2.78  &   23   &  1.33  &  1.70  &  0.71  &  1.03  &  23506.41  &  2.00 & 8  \\
HD 43691  &   36.9916$\pm$0.0350715  &   125.98$\pm$4.06  &   0.0897$^{+0.0432}_{-0.0346}$  &   91.3$\pm$26.2  &   16.7$\pm$24.9  &   2454046.7  &   ...$\pm$...  &   10.35$\pm$2.58  &   36   &  1.38  &  1.92  &  2.49  &  1.04  &  42743.17  &  2.80 & 18  \\
$\rho$ Crb  &   39.8459$\pm$0.00917865  &   65.25$\pm$2.20  &   0.0476$^{+0.0218}_{-0.0476}$  &   -62.1$\pm$42.2  &   -172.6$\pm$42.5  &   2451181.1  &   ...$\pm$...  &   0.65$\pm$2.79  &   26   &  1.00  &  1.28  &  1.09  &  1.03  &  166705.71  &  7.64 & 6  \\
HD 45652  &   43.6896$\pm$0.105825  &   35.76$\pm$2.84  &   0.4339$^{+0.0625}_{-0.0632}$  &   83.2$\pm$12.8  &   77.7$\pm$10.2  &   2453692.66  &   ...$\pm$...  &   8.38$\pm$2.20  &   45   &  0.83  &  1.04  &  0.47  &  1.01  &  12199.36  &  5.00 & 23  \\
HD 107148  &   48.6168$\pm$3.59204  &   10.01$\pm$4.15  &   0.0279$^{+0.1135}_{-0.0279}$  &   126.5$\pm$90.8  &   -50.1$\pm$88.8  &   2452799.9  &   0.003$\pm$0.001\footnotemark[2]   &   3.53$\pm$1.04  &   35   &  1.14  &  1.12  &  0.21  &  0.96  &  122900.60  &  5.60 & 6  \\
\hline
\end{longtable}
{References: 1~\citet{Bouchy2005}; 2~\citet{Melo2007};
3~\citet{Ge2006}; 4~\citet{Forveille2009}; 5~\citet{Maness2007};
6~\citet{Butler2006}; 7~\citet{Zucker2004}; 8~\citet{Johnson2006};
9~\citet{Winn2006}; 10~\citet{Winn2009}; 11~\citet{LoCurto2006};
12~\citet{Moutou2005}; 13~\citet{Pepe2004}; 14~\citet{Fischer2006};
15~\citet{Ge2006}; 16~\citet{Bonfils2007}; 17~\citet{Howard2009};
18~\citet{daSilva2006}; 19~\citet{daSilva2007}; 20~\citet{Udry2002};
21~\citet{Udry2006}; 22~\citet{Marcy2005}; 23~\citet{Mayor2004};
24~\citet{Santos2008}. Stellar radius and age estimations are
obtained from the following sources with descending priority: 1,
~\citet{Takeda2007}; 2, $nsted.ipac.caltech.edu$; 3,
$exoplanet.eu$. $\tau_{circ}$ is calculated assuming $Q^\prime_\ast=10^7$ and $Q^\prime_p=10^7$}


\end{landscape}
\twocolumn

\bsp

\label{lastpage}

\end{document}